\documentclass[amsmath,prb,twocolumn,superscriptaddress]{revtex4}
\usepackage{amsmath}
\usepackage{amssymb}
\usepackage{graphics}
\begin{document}
\title{Green's functions perspective on nonequilibrium thermodynamics of open quantum systems strongly coupled to baths}
\author{Nicolas Bergmann}
\affiliation{Department of Chemistry, Technical University of Munich, D-85748 Garching, Germany}
\author{Michael Galperin}
\email{migalperin@ucsd.edu}
\affiliation{Department of Chemistry \& Biochemistry, University of California San Diego, La Jolla, CA 92093, United States}

\begin{abstract}
We give nonequilibrium Green's function (NEGF) perspective on thermodynamics formulations for open
quantum systems strongly coupled to baths. Scattering approach implying thermodynamic consideration
of a super-system (system plus baths) weakly coupled to external super-baths is compared with
consideration of thermodynamics of a system strongly coupled to its baths. 
We analyze both approaches from the NEGF perspective and argue that the latter yields
a possibility of thermodynamic formulation consistent with dynamical (quantum transport) description. 
\end{abstract}

\maketitle


\section{Introduction}\label{intro}
Tremendous progress in experimental techniques in the last decade resulted in miniaturization
of devices for energy storage and conversion making use of quantum effects possible.
One of such experimental developments is study of thermoelectric effects in nanoscale
single atom and single molecule junctions~\cite{reddy_thermoelectricity_2007,lee_heat_2013,kim_electrostatic_2014,cui_quantized_2017,cui_thermal_2019}. 
Such quantum devices are characterized by efficiency of their performance~\cite{zotti_heat_2014,cui_perspective:_2017,cui_peltier_2018},
and traditional characteristics (such as, e.g., figure of merit)  
taken from studies on macroscopic equilibrium thermoelectric systems are often utilized.
Clearly, macroscopic thermodynamics underlying such characteristics is not applicable at nanoscale.
Meaningful description of efficiency in nanoscale junctions requires
corresponding development of quantum nonequilibrium thermodynamic theory.
Moreover, in junctions with molecules chemisorbed on (at least one of the) macroscopic contacts
thermodynamic theory should account for non-negligible (strong) system-baths couplings.

Significant theoretical effort was undertaken to formulate nanoscale thermodynamics
at strong system-bath coupling for both classical~\cite{seifert_first_2016,jarzynski_stochastic_2017} and quantum~\cite{carrega_energy_2016,katz_quantum_2016,hsiang_quantum_2018} systems.
Arguably, there are two main approaches to the problem. 
First approach complements physical system strongly coupled to its baths with set of additional 
super-baths and implements standard methods in consideration of super-system (system plus baths) 
weakly coupled to its super-baths.
System thermodynamics is defined as a difference between thermodynamic characterization of 
such super-system and that of set of free baths weakly coupled to corresponding super-baths.
Second approach builds thermodynamic description for the physical system,
i.e. system strongly coupled to its baths.
In addition to developments of thermodynamic formulations, an interesting widely debated question is
possibility of thermodynamics being consistent with underlying system dynamics~\cite{tannor_interplay_1999,kosloff_quantum_2013}.

Here, we consider a generic model of molecular junction with non-negligible (strong) molecule-contacts
couplings. We utilize nonequilibrium Green's function (NEGF) to describe dynamics of the system
and discuss compatibility of the dynamic consideration with several suggestions for thermodynamic
characterization of such systems available in the literature.
We present general NEGF formulations (beyond usually assumed slow driving) 
for thermodynamic characteristics of the system and
argue that difficulties of Green's function based analysis of supersystem-superbath thermodynamic
treatments are caused by incompatibility of basic assumptions in the two theories.
System-bath thermodynamic formulations are found to be compatible with NEGF dynamics.
Structure of the paper is the following. Section~\ref{model} introduces model
and presents basics of the dynamical NEGF treatment. Thermodynamic NEGF based formulations
are presented in Section~\ref{supersys} for supersystem-superbath and in Section~\ref{sys}
for system-bath considerations. Section~\ref{conclude} summarizes our findings. 


\section{Dynamical consideration}\label{model}
We consider an open non-interacting nonequilibrium quantum system $S$
(e.g., molecule with its electronic structure calculated using DFT) 
strongly coupled to its baths $\{B\}$ 
(e.g., metallic contacts in the junction).
The system is subjected to an arbitrary external driving.
Hamiltonian of the model is
\begin{equation}
 \label{H}
 \hat H(t) = \hat H_S(t) + \sum_{B}\bigg(\hat H_B + \hat V_{SB}(t)\bigg)
\end{equation}
where $\hat H_S(t)$ and $\hat H_B$ are Hamiltonians of the system 
and bath $B$, respectively; $\hat V_{SB}(t)$ describes coupling 
(electron transfer) between system $S$ and bath $B$.
Explicit expressions are
\begin{align}
\label{HS}
&\hat H_S(t) = 
\sum_{m_1,m_2\in S} H_{m_1m_2}^{(S)}(t) \hat d_{m_1}^\dagger\hat d_{m_2},
\\
&\hat H_B =
\sum_{k\in B} \varepsilon_k\hat c_k^\dagger\hat c_k,
\\
\label{VSB}
&\hat V_{SB}(t) = 
\sum_{m\in S}\sum_{k\in B}\bigg( V_{mk}(t) \hat d_m^\dagger\hat c_k + H.c.\bigg),
\end{align}
where $\hat d_m^\dagger$ ($\hat d_m$) and $\hat c_k^\dagger$ ($\hat c_k$)
create (annihilate) electron in orbital $m$ of the molecule and 
single-particle state $k$ of a contact, respectively.

For such a non-interacting model one can easily simulate exact
projections of the single-electron Green's function
\begin{equation}
\label{defG}
 G_{n_1n_2}(\tau_1,\tau_2) \equiv 
-i \langle T_c\, \hat a_{n_1}(\tau_1)\,\hat a_{n_2}^\dagger(\tau_2)\rangle
\end{equation}
Here, $n_i$ are indices for single-particle state either on the molecule
or in the baths, i.e. $\hat a_n$ is either $\hat d_m$ or $\hat c_k$. 

Dynamical (quantum transport) consideration defines particle, $I_B$, and 
energy, $J_B$, fluxes at $S-B$ interface as (minus) rates of change of,
respectively, particles and energy in the bath $B$.
Exact expressions for the fluxes in terms of 
single particle Green's functions are obtained 
following Jauho-Wingreen-Meir~\cite{jauho_time-dependent_1994} and 
similar~\cite{mukamel_flux-conserving_2019} derivations
\begin{widetext}
\begin{align}
\label{IB}
I_B(t) &\equiv 
-\sum_{k\in B} \frac{d}{dt}\big\langle \hat c_k^\dagger(t)\hat c_k(t)\big\rangle
= -\mbox{Tr}\bigg[ \hat N_B\, \frac{d}{dt}\hat\rho(t) \bigg]
=\sum_{k\in B}\mbox{Tr}_S\big[ \mathbf{I}_k^{(+)}(t) - \mathbf{I}_k^{(-)}(t) \big]
\\
\label{JB}
J_B(t) &\equiv 
-\sum_{k\in B} \varepsilon_k \frac{d}{dt}\big\langle \hat c_k^\dagger(t)\hat c_k(t)\big\rangle
= - \mbox{Tr}\bigg[ \hat H_B\,\frac{d}{dt}\hat\rho(t) \bigg]
= \sum_{k\in B}\varepsilon_k\,\mbox{Tr}_S\big[ \mathbf{I}_k^{(+)}(t) - \mathbf{I}_k^{(-)}(t) \big]
\end{align}
Here, $\hat N_B\equiv \sum_{k\in B}\hat c_k^\dagger\hat c_k$
is the operator of particle number in bath $B$,
$\hat \rho(t)$ is the total (system plus baths) density operator,
$\mbox{Tr}[\ldots]$ and $\mbox{Tr}_S[\ldots]$ are traces over 
total (system plus baths) and system (molecular) degrees of freedom,
$\mathbf{I}_k^{(+)/(-)}$ are matrices in subspace $S$ representing $k$-resolved
in-/out-scattering particle fluxes at the $S-B$ interface
\begin{align}
\label{Ik+}
\big[\mathbf{I}_k^{(+)}\big]_{m_1m_2} &= 2\,\mbox{Re}\int_{-\infty}^{t} dt'\,\big(
V_{m_1k}(t)\, g_k^{<}(t-t')\, V_{km_2}(t')\, G_{m_2m_1}^{>}(t',t) \big)
\\
\label{Ik-}
\big[\mathbf{I}_k^{(-)}\big]_{m_1m_2} &= 2\,\mbox{Re}\int_{-\infty}^{t} dt'\,\big(
V_{m_1k}(t)\, g_k^{>}(t-t')\, V_{km_2}(t')\, G_{m_2m_1}^{<}(t',t) \big)
\end{align}
\end{widetext}
$G_{m_2m_1}^{\lessgtr}$ are lesser/greater projections of the molecular
space single particle Green's function (\ref{defG}),
and $g_k(\tau_1,\tau_2)\equiv
-i\langle T_c\,\hat c_k(\tau_1)\,\hat c_k^\dagger(\tau_2)\rangle$
is Green's function of free electron in bath $B$.
Note, definition (\ref{JB}) assumes $\langle \hat H_B \rangle$
to be energy of bath $B$, so that dynamical approach sets
\begin{equation}
 \label{ES}
 E_S(t) = \bigg\langle \hat H_S(t) + \sum_B \hat V_{SB}(t) \bigg\rangle
\end{equation}
as energy of the system.


\section{Supersystem weakly coupled to superbaths}\label{supersys}
At equilibrium, thermodynamics of the system strongly coupled to its bath (one bath, $B$, is enough at equilibrium) is modeled
as difference in thermodynamic description (difference of grand potentials) 
of supersystem (system plus bath) weakly coupled to superbath (additional external bath) and
bath weakly coupled to the superbath.  The approach was pioneered in 
Refs.~\onlinecite{hanggi_finite_2008,ingold_specific_2009}. It allows to utilize standard (weakly coupled) 
thermodynamic description to derive grand potential, entropy and energy of the system 
as~\cite{bruch_quantum_2016}
\begin{align}
\Omega_{S}^{eq} &= \frac{1}{\beta_B}\int\frac{dE}{2\pi}\,\mathcal{A}_B(E)\ln[1-f_B(E)]
\\
\label{Seq}
S^{eq} &\equiv -\frac{\partial \Omega_S^{eq}}{\partial T_B} = \int\frac{dE}{2\pi}\, \mathcal{A}_B(E)\,\sigma_B(E)
\\
\label{ESeq}
E_S^{eq} &\equiv \Omega_S^{eq} +\mu_B N_B^{eq} +\frac{1}{\beta_B} S^{eq} =
\Omega_S^{eq} -\mu_B\frac{\partial \Omega_S^{eq}}{\partial \mu_B} +\frac{1}{\beta_B} S^{eq} 
\nonumber \\ &= \int\frac{dE}{2\pi}\, E\,\mathcal{A}_B(E)\, f_B(E)
\end{align}
Here, $\beta_B=1/k_B T_B$, $\sigma_B(E)$ is the energy-resolved Shannon entropy
\begin{equation}
\label{sigmaB}
 \sigma_B(E) = -\bigg(f_B(E)\ln f_B(E) + [1-f_B(E)]\ln[1-f_B(E)]\bigg)
\end{equation}
and $\mathcal{A}_B(E)$ is the renormalized spectral function
\begin{equation}
 \mathcal{A}_B(E) = A(E) -2\,\mbox{Im}\sum_{k\in B}\bigg[ G^r_{kk}(E) - g^r_k(E) \bigg]
 \end{equation}
 with 
 \begin{equation}
 A(E) =  -2\,\mbox{Im}\sum_{m\in S}\, G_{mm}^r(E)
\end{equation}
being the usual spectral function of the system.

At nonequilibrium, expressions (\ref{Seq}) and (\ref{ESeq}) are used as templates for {\em ad hoc}
formulations of energy and entropy by substituting spectral functions and/or Fermi distributions
with their nonequilibrium analogs at slow driving~\cite{esposito_quantum_2015,bruch_quantum_2016,haughian_quantum_2018}.
Expressions for system characteristics at slow driving are 
obtained employing gradient expansion~\cite{esposito_nature_2015,haug_quantum_2008}.
More consistent approaches to nonequilibrium  reformulate equilibrium consideration of Refs.~\onlinecite{hanggi_finite_2008,ingold_specific_2009}  in the basis
of scattering states~\cite{bruch_landauer-buttiker_2018}. 
In this formulation superbaths provide thermal distributions of the scattering states.
Parametric dependence of scattering matrix on time developed for adiabatic quantum pumps in 
Ref.~\onlinecite{moskalets_adiabatic_2004} is utilized to obtain nonequilibrium system behavior at
slow driving. It was shown within such thermodynamic considerations~\cite{bruch_quantum_2016,ludovico_dynamical_2014,ludovico_periodic_2016,ludovico_dynamics_2016} 
that consistent (dynamic-to-thermodynamic) description can be obtained within the wide-band 
approximation (WBA) and for driving confined to the system Hamiltonian $\hat H_S(t)$,
if energy of the system is taken as
\begin{equation}
\label{ESalt}
 E_S(t) = \bigg\langle \hat H_S(t) + \frac{1}{2}\sum_B \hat V_{SB}(t) \bigg\rangle
\end{equation}
Similar separation of the total Hamiltonian is assumed in recent density matrix based approaches~\cite{dou_universal_2018,alex2019transport,oz_evaluation_2019,dou2020universal}.
Extension of the formulation to account for driving in
the system-bath coupling was claimed\footnote{We note that term introduced in Ref.~\onlinecite{haughian_quantum_2018} to account for driving in the system-bath coupling
-- work done by the system-bath coupling $\dot W_B(t)$ -- is an artifact of inconsistent treatment:
the term can be derived as surface term where limit of wide band is taken first 
while limit of energy going to infinity second. Physically relevant order of taking the limits is the opposite.
In this case the term $\dot W_B(t)$ is identically zero.} in Ref.~\onlinecite{haughian_quantum_2018}.
Note that definition (\ref{ESalt}) deviates from the dynamical definition (\ref{ES}).

Before proceeding to Green's function based analysis we want to stress several points.
First, definition (\ref{ESalt}) modifies energies of the baths adding half of system-bath coupling
into the bath's energy. This addition induces mixing between baths making 
full counting statistics formulation impossible. Thus, it is natural that definition (\ref{ESalt})
fails to describe energy fluctuations~\cite{ochoa_energy_2016}.
Second, simple single particle scattering formulation is only possible for noninteracting systems
and adiabatically slow driving, when scattering channels are independent of each other.
Indeed, scattering theory yields the famous Landauer-B{\" u}ttiker formalism applicable
in description of steady-states in noninteracting systems. Finite driving and/or presence of interactions
requires more elaborated description.
Third, consistent thermodynamic description employing definition (\ref{ESalt}) was only possible 
in the wide-band approximation (WBA) where renormalization of the spectral function is dropped,
i.e. $\mathcal{A}(E)=A(E)$. As we show below, extension of the formulation beyond the WBA is impossible
when (\ref{ESalt}) is taken as energy of the system.

We now turn to NEGF analysis of the two definitions for system energy, Eqs.~(\ref{ES}) and (\ref{ESalt}),
with the goal to establish their consistency with the expected limiting (equilibrium) expression, Eq.~(\ref{ESeq}),
as obtained from general result for a noninteracting system (\ref{H})-(\ref{VSB}) 
under arbitrary driving and beyond wide-band approximation.
To do so we are going to express contributions to the total energy, i.e. averages of terms in the total 
Hamiltonian (\ref{H}), in terms of Green's functions utilizing Wigner representation in time variables
\begin{equation}
\label{Wigner}
\begin{split}
 F(t;s) &=F(t_1,t_2)
 \\
 F(t;E) &= \int ds e^{-iEs} F(t;s)
 \end{split}
\end{equation}
Here, $t=(t_1+t_2)/2$ and $s=t_1-t_2$.

First, it is straightforward to see that
\begin{equation}
\label{HSW}
\begin{split}
\big\langle \hat H_S(t) \big\rangle
&= -i \sum_{m_1,m_2\in S} H^{(S)}_{m_1m_2}(t)\, G^{<}_{m_2m_1}(t,t)
\\
&= -i\int\frac{dE}{2\pi}\mbox{Tr}_S\big[ \mathbf{H}^{(S)}\, \mathbf{G}^{<}(t;E)\big]
\end{split}
\end{equation}

Second, for system-baths coupling we get
\begin{equation}
\label{VSBW}
\begin{split}
& \sum_B \big\langle \hat V_{SB}(t) \big\rangle
= 2\,\mbox{Im}\sum_B\sum_{m\in S}\sum_{k\in B} V_{mk}(t)\, G_{km}^{<}(t,t)
\\ &=
2\,\mbox{Im}\int_{-\infty}^{+\infty} dt'\, \mbox{Tr}_S\big[ \Sigma^{<}(t,t')\, \mathbf{G}^a(t',t)
+\Sigma^r(t,t')\, \mathbf{G}^{<}(t',t)\big]
\\ &=
2\,\mbox{Im}\,\mbox{Tr}_S\bigg[\bigg(i\frac{\partial \mathbf{G}^{<}(t,t')}{\partial t}\bigg)_{t=t'} 
- \mathbf{H}^{(S)}(t)\,\mathbf{G}^{<}(t,t) \bigg]
\\ &
\equiv -2\,i\int\frac{dE}{2\pi}\, E\,\mbox{Tr}_S\big[\mathbf{G}^{<}(t;E)\big] - 2\big\langle \hat H_S(t)\big\rangle
\end{split}
\end{equation}
where transition from first to second line uses integral form of Dyson equation for $G_{km}^{<}(t,t)$,
third line is obtained employs differential from of left side Dyson equation for $G^{<}_{m_1m_2}(t,t)$
together with assumption of non-interacting character of the system, i.e. 
$\Sigma(\tau,\tau')=\sum_B \Sigma_B(\tau,\tau')$,
and last line is obtained by using Wigner representation (\ref{Wigner}) for the first term and 
by using Eq.~(\ref{HSW}) for the second term.

Third, for baths contributions to the total energy one has
\begin{equation}
\label{HBW}
\begin{split}
&\sum_B \big\langle\hat H_B\big\rangle 
= -i\sum_B\sum_{k\in B} \varepsilon_k G^{<}_{kk}(t,t)
\\ &
=\sum_B\sum_{k\in B}\mbox{Im}\bigg[\bigg(i\frac{\partial G_{kk}^{<}(t,t')}{\partial t}\bigg)_{t=t'}
-\sum_{m\in S} V_{km}(t)\, G^{<}_{mk}(t,t) \bigg]
\\ &
= \mbox{Im}\bigg[ \sum_B\sum_{k\in B}\bigg(i\frac{\partial G_{kk}^{<}(t,t')}{\partial t}\bigg)_{t=t'}
-\sum_{m\in S}\bigg(i\frac{\partial G_{mm}^{<}(t,t')}{\partial t}\bigg)_{t=t'}
\\ &\qquad\qquad\qquad\qquad\qquad
+\sum_{m_1,m_2\in S} H^{(S)}_{m_1m_2}(t)\, G^{<}_{m_2m_1}(t,t) \bigg]
\\ & =
-i \int\frac{dE}{2\pi}\, E\bigg( \mbox{Tr}_B\big[ \mathbf{G}^{<}(t;E)\big] 
- \mbox{Tr}_S\big[\mathbf{G}^{<}(t; E)\big] \bigg)
+\big\langle \hat H_S(t) \big\rangle
\end{split}
\end{equation}
Here, transitions from first to second and from second to third and fourth lines
utilize differential forms of left side Dyson equations for $G_{kk}^{<}(t,t)$ and $G_{mm}^{<}(t,t)$, respectively.
As previously, last line is obtained by using Wigner representation (\ref{Wigner}) for the first and second terms 
and  by using Eq.~(\ref{HSW}) for the last term.
Similarly, for free baths evolution one has
\begin{equation}
\label{HB0W}
\begin{split}
\sum_B \big\langle\hat H_B\big\rangle_0 
&= -i\sum_B\sum_{k\in B} \varepsilon_k\, g^{<}_{k}(t,t)
\\
&=\sum_B\sum_{k\in B}\mbox{Im}\bigg[\bigg(i\frac{\partial g_{k}^{<}(t,t')}{\partial t}\bigg)_{t=t'}\bigg]
\\ &
= -i\int\frac{dE}{2\pi}\, E\, \mbox{Tr}_B\, \mathbf{g}^{<}(E)
\end{split}
\end{equation}
where $\mathbf{g}^{<}(E)$ does not contain dependence on $t$ due to absence of driving in baths.

We note that contrary to previous considerations expressions (\ref{HSW})-(\ref{HB0W}) 
are not limited to slow driving -- for non-interacting model (\ref{H})-(\ref{VSB}) they are exact.
Eqs.~(\ref{HSW}) and (\ref{VSBW}) show that dynamical definition (\ref{ES}) does not yield
expected within the approach equilibrium behavior (\ref{ESeq}), while scattering theory based
suggestion, Eq.~(\ref{ESalt}), leads to
\begin{equation}
\bigg\langle\hat H_S(t) + \frac{1}{2}\sum_B \hat V_{SB}(t)\bigg\rangle =
-i\int\frac{dE}{2\pi}\, E\, \mbox{Tr}_S\big[G^{<}(t;E)\big]
\end{equation}
At equilibrium, this expression yields result similar to (\ref{ESeq}) but with $\mathcal{A}(E)$
substituted with $A(E)$, i.e. one gets the form of correct limiting expression in the wide band approximation (WBA). 
It is clear from the derivation above that generalization beyond WBA is not possible when
using (\ref{ESalt}) as definition for system energy.

To get the expected equilibrium behavior, Eq.~(\ref{ESeq}), one has to assume
\begin{equation}
\label{ESalt2}
 E_S(t) = \bigg\langle \hat H_S(t) + \sum_B\bigg( \hat H_B + \hat V_{SB}(t) \bigg)\bigg\rangle
  - \bigg\langle \sum_B \hat H_B \bigg\rangle_0
\end{equation}
as expression for system energy. Here, $\langle\ldots\rangle = \mbox{Tr}[\ldots \hat\rho(t)]$ and
$\langle\ldots\rangle_0 = \mbox{Tr}[\ldots \hat\rho_0]$ with $\hat \rho_0$ being density operator of free
decoupled system and baths evolution.
Indeed, substituting (\ref{HSW})-(\ref{HB0W}) into (\ref{ESalt2}) leads to
\begin{equation}
 E_S(t) = -i\int\frac{dE}{2\pi}\, E \bigg( \mbox{Tr}\big[\mathbf{G}^{<}(t;E)\big] - \mbox{Tr}_B\big[\mathbf{g}^{<}(E) \big]\bigg)
\end{equation}
which yields the expected equilibrium behavior. 

We note that expression (\ref{ESalt2}) is very logical
in a sense that it follows philosophy of defining system characteristics as difference between
those of supersystem and free baths. At the same time, it reveals basic incompatibility
between supersystem weakly coupled to superbaths thermodynamic approach
and standard NEGF dynamical formulation. Lack of superbaths concept in the latter does not
allow to meaningfully introduce heat in any attempt of combining the two descriptions.

\section{System strongly coupled to baths}\label{sys}
A variant of thermodynamic formulation for system strongly coupled to its baths was proposed in
Refs.~\onlinecite{esposito_entropy_2010,ptaszynski_entropy_2019}. 
As expected, in absence of superbaths definition of system energy 
\begin{equation}
 E_S(t) = \bigg\langle \hat H_S(t) + \sum_B \hat V_{SB}(t) \bigg\rangle
 - \bigg\langle \hat H_S(t) + \sum_B \hat V_{SB}(t) \bigg\rangle_0
\end{equation}
and expression for energy flux are consistent with dynamical NEGF results - 
Eq.~(\ref{ES}) and (\ref{JB}), respectively.
Similar to the supersystem-superbath thermodynamics,
system-bath formulation is also based on a set of {\em ad hoc} assumptions.
In particular, Ref.~\onlinecite{esposito_entropy_2010} assumes entropy of the system strongly
coupled to its baths to be given by the Shannon entropy
\begin{equation}
 \label{Shannon}
 S(t) \equiv -\mbox{Tr}_S\big[ \hat \rho_S(t)\,\ln\hat \rho_S(t)\big]
\end{equation}
where $\hat \rho_S(t) = \mbox{Tr}_{B}[\hat \rho(t)]$ is the many-body density operator of the system.
Below we show how entropy, Eq.~(\ref{Shannon}), and the second law of themodynamics
can be expressed in terms of Green's functions.

First, we note that for quadratic Hamiltonian (\ref{H})-(\ref{VSB}) 
the Wick's theorem holds for the whole universe (system plus baths) or any of its parts. 
This means that corresponding many-body density operator
should have a Gaussian form. In particular, system density operator has the form
\begin{equation}
 \label{rhoS}
 \hat \rho_S(t) = 
\frac{1}{Z_S(t)} \exp\bigg(-\sum_{m_1,m_2\in S}h^{(S)}_{m_1m_2}(t)\,\hat d_{m_1}^\dagger\,\hat d_{m_2}\bigg)
\end{equation}
where $Z_S(t)$ is a normalization constant.
The form (\ref{rhoS}) is mathematically similar to equilibrium case, 
so that for fixed $t$ standard tools of equilibrium path integral consideration can be applied.
In particular, we can consider the form (\ref{rhoS}) as 
an equilibrium density matrix with `effective Hamiltonian'
$\sum_{m_1,m_2\in S}h^{(S)}_{m_1m_2}(t)\hat d_{m_1}^\dagger\hat d_{m_2}$
and inverse temperature $\beta_S=1$.

Using results of equilibrium consideration for non-interacting 
Hamiltonians~\cite{negele_quantum_1988} one gets
\begin{align}
 \label{ZSt}
 Z_S(t) &= \mbox{det}\big[i\mathbf{G}^>(t,t)\big]^{-1}
 \\
 \label{ht}  
 e^{-\mathbf{h}^{(S)}(t)} &= 
\big(i\mathbf{G}^{>}(t,t)\big)^{-1}\big(-i\mathbf{G}^{<}(t,t)\big)
\\ &
=
\big(-i\mathbf{G}^{<}(t,t)\big)\big(i\mathbf{G}^{>}(t,t)\big)^{-1}
\nonumber
\end{align}
Thus,
\begin{equation}
\label{rhoSG}
\begin{split}
 \mathbf{\rho}_S(t) &= \frac{1}{Z_S(t)} e^{-\mathbf{h}^{(S)}(t)}
\\ &= \mbox{det}\big[i\mathbf{G}^{>}(t,t)\big]
\big(-i\mathbf{G}^{<}(t,t)\big)\big(i\mathbf{G}^{>}(t,t)\big)^{-1}
\\
& = \mbox{det}\big[i\mathbf{G}^{>}(t,t)\big]
\big(i\mathbf{G}^{>}(t,t)\big)^{-1}\big(-i\mathbf{G}^{<}(t,t)\big)
\end{split}
\end{equation}
Here, $\mathbf{G}^\gtrless(t,t)h$ are matrices of greater/lesser projection of Green's function (\ref{defG})
in the system subspace of the problem, $\mathbf{\rho}_S(t)$ is the system density matrix (representation
of operator $\hat \rho_S(t)$ in the single-particle basis of $S$).
Similar expressions were derived in Refs.~\onlinecite{chung_density-matrix_2001,cheong_many-body_2004,peschel_reduced_2009,sharma_landauer_2015}.

Using (\ref{rhoSG}) in (\ref{Shannon}) and employing
$\ln\mbox{det}\, \mathbf{M} = \mbox{Tr}\,\ln\mathbf{M}$ ($\mathbf{M}$ is a matrix) leads to
\begin{widetext}
\begin{equation}
\label{SG}
 S(t) = 
-\mbox{Tr}_S\bigg[-i\mathbf{G}^{<}(t,t)\,\ln\big(-i\mathbf{G}^{<}(t,t)\big)\bigg]
-\mbox{Tr}_S\bigg[i\mathbf{G}^{>}(t,t)\,\ln\big(i\mathbf{G}^{>}(t,t)\big)\bigg]
\end{equation}
Here, $-i\mathbf{G}^{<}(t,t)$ is the single-particle density matrix $\rho_S(t)$
and $i\mathbf{G}^{>}(t,t)-i\mathbf{G}^{<}(t,t)=\mathbf{I}$ ($\mathbf{I}$ is the unity matrix).
Note that (\ref{SG}) holds for any driving.

Taking time derivative of the entropy (\ref{SG}) leads to the second law of thermodynamics in the form
\begin{align}
\label{second}
 \frac{d}{dt} S(t) &= \mbox{Tr}_S\bigg[\bigg(-i\frac{d}{dt}\mathbf{G}^{<}(t,t)\bigg)\,
 \ln\frac{i\mathbf{G}^{>}(t,t)}{-i\mathbf{G}^{<}(t,t)}\bigg]
 \\ &\equiv
 \sum_B\sum_{k\in B}\mbox{Tr}_S\bigg[
\beta_B(\varepsilon_k-\mu_B)\big[\mathbf{I}_k^{(+)}(t)-\mathbf{I}_k^{(-)}(t)\big]
 + \big[\mathbf{I}_k^{(+)}(t)-\mathbf{I}_k^{(-)}(t)\big]\,  
 \ln\frac{\mathbf{G}^{>}(t,t)\, g_k^{<}(t,t)}{\mathbf{G}^{<}(t,t)\, g_k^{>}(t,t)}
 \bigg]
\nonumber \\
& = \sum_B \big[ \beta_B\, \dot Q_B(t) + \Delta_i \dot S_B(t)\big]
\nonumber
\end{align}
\end{widetext}
Here, $\dot Q_B\equiv J_B(t)-\mu_B I_B(t)$ is the heat flux expressed in terms of particle, $I_B$, 
and energy, $J_B$, fluxes at $S-B$ interface, 
whose definitions are given by the dynamical NEGF  expressions (\ref{IB}) and (\ref{JB}), respectively.
$\Delta_i \dot S(t)=\sum_B\Delta_i \dot S_B$ 
defined by the second term in the middle line of (\ref{second}) is the rate of entropy production 
which (as was discussed in Ref.~\onlinecite{esposito_entropy_2010}) may be negative. 

 
\section{Conclusion}\label{conclude}
We considered two different approaches to thermodynamic formulations for open nonequilibrium
quantum systems strongly coupled to their baths: supersystem (system plus baths) weakly
coupled to superbaths and system strongly coupled to its baths. 
In particular, the former encompasses popular scattering theory formulations of quantum thermodynamics.
We analyzed compatibility of the formulations with dynamical description of the system within 
the nonequilibrium Green's function approach.
We presented thermodynamics formulation within NEGF beyond slow driving.
Results for adiabatic driving and equilibrium can be derived from our consideration as limiting cases.
Our analysis shows that supersystem-superbath
formulations are based on set of assumptions which are incompatible with basics of the dynamical
NEGF formulation. In particular, this is the cause for difference in definition of energy flux as
accepted in the two approaches. 
At the same time, the system-bath formulation is consistent
with NEGF, and definitions of energy fluxes are equivalent in this thermodynamic formulation
to those of dynamic NEGF description. For the system-bath formulation we present expressions
for entropy and entropy production in terms of Green's functions.
It is interesting to note that while supersystem-superbath formulations postulate energy resolved
Shannon-like expression for entropy of the system, system-bath approach assumes
entropy of the system to be given by Shannon expression constructed from system characteristics 
integrated in energy. We note that both expressions for system entropy, nonequilibrium analog 
of Eq.~(\ref{Seq}) and Eq.~(\ref{SG}), 
are {\em ad hoc} formulations, and possibility of construction of energy-resolved formulation consistent
with dynamical NEGF description is still an open question.

\begin{acknowledgments}
We acknowledge helpful discussions with Abraham Nitzan.
This material is based upon work supported by the National Science Foundation under CHE-1565939.
NB thanks University of California San Diego for financial support and hospitality during his visit.
\end{acknowledgments}

%

\end{document}